\documentstyle[preprint,aps]{revtex}

\begin{document}
\draft
\title{Far-infrared edge modes in quantum dots}
\author{E. Lipparini$^{1}$\cite{perm}, N. Barber\'an$^1$,
M. Barranco$^1$, M. Pi$^1$, and Ll. Serra$^2$.}
%
%
\address{$^1$Departament d'Estructura i Constituents de la Mat\`eria,
Facultat de F\'{\i}sica, \\
Universitat de Barcelona, E-08028 Barcelona, Spain}
\address{$^2$Departament de F\'{\i}sica, Facultat de Ci\`encies,\\
Universitat de les Illes Balears, E-07071 Palma de Mallorca, Spain}
\date{\today}

\maketitle

\begin{abstract}

We have investigated  edge modes of different
multipolarity sustained by quantum dots submitted to external
magnetic fields. We present a microscopic
description based on a variational solution
of the equation of motion for any axially symmetric confining
potential and multipole mode. Numerical results  for
dots with different number of electrons
whose ground-state  is described within a local Current
Density Functional Theory are discussed. Two sum rules, which are
exact within this theory, are derived.
 In the limit of a large neutral dot at B=0, we have shown that
the classical hydrodynamic dispersion law for edge waves
$\omega$(q) $\sim \, \sqrt{q\, \ln \,(q_0/q)}$  holds when
quantum and finite size effects are taken into account.

\end{abstract}


\narrowtext
\section*{}

\section{Introduction}

Collective excitations induced in finite  fermion
systems by external probes have been extensively studied in the last
 years. Particular effort has been devoted to the understanding of the
giant dipole
resonance in nuclei \cite{lip}, and of the plasmon mode in metallic
clusters \cite{bra,deh,roc}. Recently, a strong collective state has
also been
observed in quantum dot structures \cite{sik,dem}. These
collective modes have all in common the feature of being L=1, S=0
excited
states induced by a dipolar external radiation which is the dominant
component of the electromagnetic field when its wavelength is much
larger than the size
of the system. Dipole modes correspond to oscillations of protons
against  neutrons in nuclei, and of electrons against the
positive background in metal clusters and dot structures. Whereas in
nuclei the restoring force of the collective motion is  the
symmetry potential which acts differently on protons than on  neutrons,
 in
metal clusters and dots it arises from the Coulomb interaction between
ions and electrons.

Collective states with multipolarity  L$>$1 were
predicted to exist and have been observed in nuclei. At present,
we have a complete systematics of quadrupole (L=2) and octupole (L=3)
collective excitations in nuclei \cite{lip,boh,ber}.
Multipole collective states in $^3$He droplets have also been studied
 \cite{ser2,ser3,wei}, but have not been experimentally detected
so far. In metal clusters, the
predicted multipole states \cite{eka,ser} have not  been observed either.
The basic reason is the
experimental difficulties arising from the fact that free clusters are
produced and analyzed on-fly, with the added
difficulty in the case of $^3$He drops of being
electrically neutral systems.

Far-infrared absorption spectroscopy experiments on large radius
quantum dots submitted to a static external magnetic field B
have likely evidenced quadrupole excitations \cite{dem}, and
an anticrossing between the  L=1 and L=2 resonances, each of them
splitted in two branches, one with negative and another with
positive B-dispersion. We recall that for these systems, L has to be
understood as the angular momentum about an axis perpendicular to
the dot plane.

The collective spectrum of quantum dots has been addressed in different
ways. An explanation based on classical arguments can be found in
\cite{dem}. In Refs. \cite{gla,mar,san}, use has been made of an
edge-magnetoplasmon model, whereas in Refs. \cite{bro,gud} a
Hartree-RPA method has been employed, and a Hartree-Fock-RPA one in
Ref. \cite{gud1}. Multipole modes have been considered in Refs.
\cite{gla} and \cite{gud}. All these approaches  neglect
the electron correlation energy, and therefore, no microscopic
calculation of L$\geq$1 modes exists which incorporates exchange
correlation energy as well as  quantum effects arising from
finite size and shell structure of dots in a magnetic field.
A workable, yet reliable formalism that takes into account all these
effects is called for to describe these excitations. In this
paper we present one of such possible formalisms. It is based on the
equation of motion method within the framework of a local Current
Density Functional Theory (CDFT) \cite{vig}. This functional theory
is well suited to study electronic systems in  presence of a magnetic
field, and has been successfully employed  to study
ground state (g.s.) properties of quantum dots \cite{fer}.

This paper is organized as follows. We describe the equation of motion
method in Section II, as well as the basics of the strength function
and its moments (sum rules) needed to interpret the experimental and
theoretical results. In Section III we present a rigorous solution of
the dipole mode in the case of a parabolic lateral confining
potential.
The exactness of the dipole solution in this case, irrespective of the
value of the magnetic field, has been  discussed in \cite{mak,bak}.
Here, we have obtained this results in a way that yields an
explicit expresion not only for the spectrum, but also for the
eigenstates.
 In Section IV we present a variational approach to the description
of multipole excitations. We show in Section V that for neutral
large dots at B=0, these excitacions build an edge wave with
dispersion relation of the type $\omega(q) \sim \sqrt{q \ln (q_0/q)}$.
Detailed numerical results are presented in Section VI for dots with
different number of electrons, and the concluding remarks
are presented in Section VII. Finally, an appendix
contains some technical details.

\section{The equation of motion method}

We consider N electrons moving in the z$=$0 plane where they are
confined by the dot potential V$_+$(r) with r $=\sqrt{x^2+y^2}$. On
this system it may act a constant magnetic field in the z-direction
described by the vector potential $\vec{A}=\frac{1}{2}(-$y,x,0)B,
and we suppose that the system can be described
by the N-electron Pauli's Hamiltonian:
\begin{equation}
H=\sum_{i=1}^{N}\,\,\left\{\frac{1}{2m^*}\left[\vec{p}_i+\frac{e}{c}
\vec{A}(\vec{r}_i)\right]^2 + V_+(r_i)+ 
\frac{1}{2} g^* \mu^*_B \vec{\sigma}_i \cdot
\vec{B}\right\}+
\sum_{i<j}^N V_c(\mid \vec{r}_i-\vec{r}_j\mid)\,\,,
\label{eq1}
\end{equation}
where m$^*$ is the electron effective mass which together with a
dielectric constant $\epsilon$ and gyromagnetic factor g$^*$ are
characteristics of the semiconductor (for example, g$^*=-$0.44,
$\epsilon$=12.4 and m$^*$=0.067 m$_e$ in GaAs), $\mu^*_B$ is the
effective
Bohr magneton $\mu _B^*=\hbar e/(2 m^* c)$, $\vec{\sigma}$ is the
Pauli matrix vector, and V$_c$ is the electron-electron (e-e) interaction
\begin{equation}
V_c(\mid \vec{r}_i-\vec{r}_j \mid)=\frac{e^2}{\epsilon} \frac{1}{ \mid
\vec{r}_i-\vec{r}_j \mid }\,\,\,\,.
\label{eq2}
\end{equation}

Eq. (\ref{eq1}) can be rewritten as
\begin{equation}
H=\sum_{i=1}^N \left\{\frac{\vec{p}\,^2_i}{2m^*}+\frac{1}{2}
\omega_cl_{z_i}+\frac{1}{8}m^*\omega^2_cr^2_i
+ \frac{1}{2} g^* \mu^*_B B \sigma_{z_i}+
V_+(r_i)\right\}+ \sum^N_{i<j} V_c(\mid \vec{r}_i-\vec{r}_j\mid)\,\,,
\label{eq3}
\end{equation}
where $\omega_c$=eB/m$^*$c is the cyclotron frequency and l$_{z_i}$ is the
angular momentum operator about the z-axis:
\begin{equation}
l_{z}=-i\hbar \frac{\partial}{\partial \theta}\,\,\, .
\label{eq4}
\end{equation}

Given the exact g.s. $\mid$0$\rangle$ of the N-electron system, it
is possible to obtain the exact spectrum corresponding to a broad
class of
collective vibrations if one is able to find an operator
O$^+$ such that the following equation of motion is fulfilled:
\begin{equation}
[H,O^+]=\hbar\omega O^+    \,\, .
\label{eq5}
\end{equation}

The state O$^+$$\mid$0$\rangle$ has an excitation energy $\hbar
\omega$, and the g.s. fulfills O$\mid$0$\rangle$=0. As
the excited states have a well defined angular momentum, so
the operators O$^+$ must have. Consequently, one has to solve Eq.
(\ref{eq5}) for each L-value.

When a magnetic field acts perpendicularly on the dot, it causes
a splitting of the excited B=0 states into two branches of energy
$\hbar \omega_{\pm L}$, each carrying an  angular momemtum $\pm\hbar$L
 over that of the g.s. It implies that besides
Eq. (\ref{eq5}), the physically acceptable O$^+_{\pm L}$ operators have
to fulfill
\begin{equation}
[L_z,O^+_{\pm L}]=\pm \hbar\,L\,O^+_{\pm L}
\label{eq6}
\end{equation}
with L=1,2,3,... and L$_z= \sum^N_{i=1}\,\,l_{z_i}$. If $\hbar$L$_0$
is the angular momentum of $\mid$0$\rangle$, the states
$\mid \pm L \rangle \equiv$
O$^+_{\pm L}\mid 0\rangle$ have an angular momentum
$\hbar$(L$_0 \pm$ L).

To easy the formulae that otherwise would be very cumbersome, from
now on we shall be using effective atomic units, defined by $\hbar=
e^2/\epsilon=m^*=$ 1. In this units system, the length unit is the
Bohr radius a$_0$ times $\epsilon$/m$^*$,
and the energy unit is the Hartree times m$^*$/$\epsilon^2$,
 which we call respectively, a$^*_0$ and E$^*_H$. For GaAs we have
 a$^*_0 \sim$ 97.94 $\AA$ and E$^*_H\sim$11.86 meV.

It is obvious that to find these
operators is in general as difficult a task as to solve the
Schr\"{o}dinger equation corresponding to the Hamiltonian
Eq. (\ref{eq1}) for the  vibrational states,
and one is led to solve Eq. (\ref{eq5}) in an approximate way, one of
this being, for example, the Random Phase Approximation. Another
 fruitful approximation, originally proposed by Feynman
to describe density excitations of superfluid $^4$He \cite{fey},
 consists of making
an ansatz on O$^+$$\mid 0 \rangle$.
Acting upon $\mid$0$\rangle$ with Eq. (\ref{eq5}) and projecting
onto O$^+\mid$0$\rangle$ one  gets
\begin{equation}
 \omega=\frac{\langle 0\mid [O,[H,O^+]]\mid 0 \rangle}
{\langle 0 \mid [O,O^+] \mid0 \rangle} \,\, .
\label{eq7}
\end{equation}
Eqs. (\ref{eq5}) and (\ref{eq7}) are completely equivalent if
$\mid$0$\rangle$ is the
exact ground state and O$^+$ is the sought operator.  The advantage of
Eq. (\ref{eq7}) is that we may look for approximate solutions of
variational type,
 guessing O$^+$ and obtaining $\mid$0$\rangle$ within a workable, yet
accurate scheme, such as the Local Density Approximation (LDA) at
B=0, or CDFT  at B$\neq$ 0. To
find these approximate solutions is the subject of Section IV.
 We shall also see that, remarkably, Eqs. (\ref{eq1}) and
(\ref{eq7}) have exact solutions in the dipole case when the confining
potential has a parabolic form.

The excitation spectrum of the system is usually probed by different
external fields, or given the well defined angular momentum of the
excited states, by a selected L-polar component of the field. For an
excitation operator F representing it, a useful, often
experimentally accesible quantity, is the so-called strength function:
\begin{equation}
S(E)=\sum_n \mid \langle n\mid F\mid 0\rangle \mid^2 \delta(E-E_n)\,\, ,
\label{eq8}
\end{equation}
where E$_n$ and $\mid$n$\rangle$ are, respectively, the excitation energy
and the excited state, and the sum or integral in the case of
continuum spectrum extends over all excited states of the system.
Of special interest are some energy moments of the strength function
\begin{equation}
m_k=\int\,\, E^k\,\,S(E)\,\,=\,\,\sum_n E_n^k\mid
\langle n\mid F\mid 0 \rangle \mid^2
\label{eq9}
\end{equation}
which we shall call sum rules (SR). They are the m$_1$ and m$_3$
moments, which can be also written as
\begin{eqnarray}
m_1 & = & \frac{1}{2} \langle 0\mid [F^+,[H,F]]\mid 0 \rangle
\nonumber
\\
& &
\label{eq10}
\\
m_3 & = & \frac{1}{2} \langle 0\mid [[H,[H,F^+]],[H,F]]\mid 0 \rangle
\,\, .
\nonumber
\end{eqnarray}
These SR have been extensively studied in the literature \cite{lip,boh}.
 For the
present purposes it is enough to recall that, if only one excited
state is contributing to S(E), E$_3 \equiv (m_3/m_1)^{1/2}$ coincides
with the corresponding excitation energy. In a more physical
situation, whenever the strength is concentrated in a  narrow,
"high" energy region, E$_3$ is a fair approximation to the
resonant energy. We  call E$_3$ the scaling energy because
m$_3$ can also be obtained scaling the $\mid$0$\rangle$ wave function as
\begin{equation}
\mid \eta\rangle= e^{\eta[H,F]} \mid 0 \rangle   \,\, ,
\label{eq11}
\end{equation}
and then carrying out the following derivative:
\begin{equation}
m_3=\frac{1}{2} \frac{\partial^2}{\partial \eta^2}
\langle \eta\mid H \mid \eta \rangle \mid_{\eta=0} \,\, .
\label{eq12}
\end{equation}
The m$_3$ sum rule can be  obtained either from Eq. (\ref{eq10}) or
from Eq. (\ref{eq12}). The latter allows one to identify m$_3$ as the
restoring
force that causes the collective vibration, and consequently, to
identify m$_1$ with the collective mass parameter.

We want to stress that not all possible approximation schemes
 fulfill the m$_1$ and m$_3$ sum rules in the sense that
a direct evaluation of Eq. (\ref{eq9}) yields the same result as Eqs.
(\ref{eq10}). A test on the consistency of the approximations made to
get the g.s.  and excitation spectrum, is the fulfillment of
these SR, especially of m$_1$ which is very model independent.

\section{Exact solution of the dipole mode for the
parabolic confining potential}

It has been frequently argued in the literature that V$_+$(r)=
$\frac{1}{2} \omega^2_0 r^2$ can be a good approximation to the
confining potential.
 That is the case, for example, when the number of
electrons in the dot is small as compared with the number of positive
ions N$_+$ that produce the confining potential \cite{mer}.
If that approximation holds, it is easy to check that the solutions of
Eq. (\ref{eq7}) for the Hamiltonian Eq. (\ref{eq3}) and L=1 are
\begin{eqnarray}
O^+_{+} & = & \frac{1}{2} \sqrt{\frac{ \bar{\omega}}{N
}}(Q-\frac{i}{\bar{\omega}}P)
\nonumber
\\
& &
\label{eq13}
\\
O^+_{-} & = & \frac{1}{2} \sqrt{\frac{ \bar{\omega}}{N
}}(Q^+-\frac{i}{\bar{\omega}}P^+)\,\,\,,
\nonumber
\end{eqnarray}
where
\begin{eqnarray}
Q & = &\sum^N_{i=1}(x_i+iy_i)\,\,\equiv \,\, \sum^N_{i=1} q^i
\nonumber
\\
& &
\label{eq14}
\\
P & = & \sum^N_{i=1}(p_{xi}+ip_{yi})\,\, \equiv \,\, \sum^N_{i=1} p^i
\nonumber
\end{eqnarray}
and
\begin{equation}
\bar{\omega}=\sqrt{\omega^2_0+\frac{\omega_c^2}{4}}\,\,\,.
\label{eq15}
\end{equation}
The corresponding frequencies are
\begin{equation}
\omega_{\pm}=
\bar{\omega}\,\,\pm\,\,  \frac{\omega_c}{2}   \,\, ,
\label{eq16}
\end{equation}
and it is easy to verify that
\begin{equation}
[L_z, O^+_{\pm}]=\pm  O^+_{\pm} \,\, .
\label{eq17}
\end{equation}
The states O$^+_{\pm}\mid$0$\rangle$ are normalized to unity and
orthogonal, carrying an angular momentum (L$_0\pm$1).
This exact result stems from the translational invariance
of the electron-electron interaction  for which
$\left[ \sum^N_{i<j} V_c(\mid \vec{r}_i-\vec{r}_j\mid)\,, P \right]$
=0, and consequently it is independent of the e-e interaction
provided it is local \cite{bak}.

It is a simple matter to check that the states $\mid\pm$1$\rangle$
exhaust the m$_1$ and m$_3$ SR for the dipole
operator D=$\sum^N_{i=1} x_i$, and that the dipole strength is equally
distributed between them. One gets
\begin{equation}
m_1(D)=\frac{1}{2}\,\,N
\nonumber
\end{equation}
\begin{equation}
m_3(D)=\frac{1}{2}\,\,N(\bar{\omega}^2+\frac{3}{4}
\omega^2_c)
\label{eq18}
\end{equation}
\begin{equation}
\mid\langle 0 \mid D\, \mid +1 \rangle \mid^2=
\mid \langle 0\mid D\,\mid -1 \rangle \mid^2=
\frac{1}{4} \frac{ N}{ \bar{\omega}}  \,\, ,
\label{eq19}
\end{equation}
and for example,
\begin{equation}
\omega_+\mid \langle 0 \mid D\, \mid +1 \rangle \mid^2+\,\,
\omega_-\mid\langle 0
\mid D\, \mid -1 \rangle \mid^2= \frac{ N}{2}= m_1(D)
\label{eq20}
\end{equation}
It is interesting to notice that in the B=0 limit,
$\omega_+= \omega_-$= E$_3$(D) = $ \omega_0$. This result is independent of the number
of electrons  in the dot, in agreement with the generalized Kohn
theorem \cite{kohn}. The parabolicity of
the  potential is expected to break down when the number of the
electrons in the dot increases and the  electronic density extends
up to the edge of the dot.
Departure of V$_+$(r) from the parabolic law originates an
N-dependence in $ \omega_{\pm}$ and in E$_3$(D).

The exact solution to Eq. (\ref{eq5}) can  be used to obtain the static
dipole polarizability of the dot $\alpha$(D), which is twice the
$m_{-1}$(D) sum rule \cite{lip,boh}:
\begin{equation}
m_{-1}(D)=\frac{1}{ \omega_+}
\mid \langle 0\mid D \,\mid +1 \rangle\mid^2
+\frac{1}{ \omega_-} \mid\langle 0\mid D \,\mid -1 \rangle \mid^2=
\frac{N}{2  \omega^2_0}  \,\, .
\label{eq21}
\end{equation}
Consequently, in the parabolic potential approximation
the dipole polarizability is independent of the magnetic
field. We expect $\alpha$(D)  to be  B-dependent in the case
of a more general confining potential.
 Using that \cite{mer}
\begin{equation}
\omega_0^2= \frac{N_+}{ R^3}
\label{eq22a}
\end{equation}
 we get
\begin{equation}
\alpha(D)= R^3 \frac{ N}{ N_+}    \,\,  .
\label{eq22}
\end{equation}
The R-dependence could have been anticipated from a dimensional analysis.

Let us finally discuss the nature of the dipole modes
with regard to their geometrical shape. If we
consider the transition density associated with the excited
states $\mid\pm$1$\rangle$
\begin{equation}
\rho_{tr}(\vec{r})= \langle 0\mid \hat{\rho}\,O^+_{\pm}
\mid 0\rangle = \langle 0\mid
[\hat{\rho}, O^+_{\pm}]
\mid
0\rangle
\,\,\,,
\label{eq23}
\end{equation}
where $\hat{\rho}$ is the N-electron density operator
\begin{equation}
\hat{\rho}= \sum^N_{i=1}\,\, \delta(\vec{r}-\vec{r}_i)\,\,\, ,
\label{eq24}
\end{equation}
it is easy to obtain
\begin{equation}
\rho_{tr}(\vec{r}) \propto \rho'(r) e^{\pm i\theta}\,\,\,,
\label{eq25}
\end{equation}
where $\rho(r)$ is the g.s. electronic density and the prime denotes the
r-derivative.  This transition density is characteristic of an
edge excitation.

\section{A variational solution for  general axially symmetric
potentials}

When the confining potential is no longer parabolic, irrespective
of the value of the magnetic field and of the multipolarity of the
mode, the equation of motion (\ref{eq5}) cannot be solved exactly,
and we have resorted to an approximate
method based on Eq. (\ref{eq7}). In the dipole case, the natural
guess for the O$^+_{\pm}$ operators is:
\begin{eqnarray}
O^+_+ & = & a_+(Q-ib_+P)
\nonumber
\\
& &
\label{eq26}
\\
O^+_- & = & a_-(Q^+-ib_-P^+) \,\,\ ,
\nonumber
\end{eqnarray}
where b$_{\pm}$ are  variational parameters and
 a$_{\pm}$ have to be determined from the normalization condition.
 This process can indeed be carried out, and one gets the
dipole spectrum of the system.

At this point, we have considered it more convenient to generalize
the Q and P operators in such a way that the calculation can be
done for any L-value. To this end, we have first taken the following
Q$_L$
\begin{equation}
Q_L=\sum^N_{i=1} r^L_i e^{iL\theta_i}\,\,\equiv \sum^N_{i=1}
q^i_L\,\,\,.
\label{eq27}
\end{equation}
This choice is inspired in  that (kr)$^L$ e$^{iL\theta}$ is the small
k-expansion of the function J$_L$(kr) e$^{iL\theta}$, which is the
restriction to
the z=0 plane of the general solution of the Laplace equation in
cylindrical coordinates. J$_L$ is the L-Bessel function of the
first kind \cite{gra}.

Next, we have taken as partner of Q$_L$ in the O$^+$ expression the
following operator:
\begin{equation}
R_L=\sum^N_{i=1} q_{L-1}p   \,\, ,
\label{eq28}
\end{equation}
where the $i$-particle index is implicit in the particle and momentum
coordinates. This choice is again guided by the exact dipole case,
since the conmutator $[H,Q_L]$ yields a combination of Q$_L$ and
R$_L$ which reduces to the one we have found in Section III when L=1.

We have thus considered as O$^+_{\pm L}$ operators the following
combinations:
\begin{eqnarray}
O^+_{+L} & = & a_+(Q_L-ib_+LR_L)
\nonumber
\\
& &
\label{eq29}
\\
O^+_{-L} & = & a_-(Q^+_L-ib_-LR^+_L)
\nonumber
\end{eqnarray}
Eqs. (\ref{eq29})  reduce to Eqs. (\ref{eq26}) for L=1, and the
proposed operators fulfill Eq. (\ref{eq6}). These modes are edge modes
whose transition densities read
\begin{equation}
\rho^{\pm L}_{tr}(\vec{r})\,=\,
\langle 0\mid
[\hat{\rho}, O^+_{\pm L}]\mid 0\rangle\, \propto \,r^{L-1}
\rho'(r) e^{\pm iL\theta} \,\, .
\label{eq30}
\end{equation}

To minimize Eq. (\ref{eq7}),
it is necessary to have a description of the g.s.
$\mid$0$\rangle$. We have taken for it the Kohn-Sham sate built from
single particle (s.p.) wave functions obtained in the framework of
the CDFT of \cite{fer}. We refer the reader to that work
for the details. For the present purposes, it
is enough to recall that the s.p. wave functions
$\phi_{nl\sigma}(r,\theta)$ are separable in $r$ and $\theta$
variables as
\begin{equation}
\phi_{nl\sigma}(r,\theta)= u_{nl\sigma}(r) e^{-il\theta}  \,\, ,
\label{eq31}
\end{equation}
with $l=0,\pm 1,\pm 2,...$ being the orbital angular momentum of the
s.p. state.
Upon minimization one gets:
\begin{equation}
\omega_{\pm L}=\frac{1\pm b_{\pm}( L\,\omega_c+4L^2\,\tilde{\omega}_1)+
b^2_{\pm}\,(\tilde{\omega}_3^2+L^3\, \omega_c
\tilde{\omega}_1)}{2\,b_{\pm} \pm 2\,b^2_{\pm}\,L^2\,\tilde{\omega}_1}
\label{eq33}
\end{equation}
\begin{eqnarray}
b_{\pm} & = & \frac{\sqrt{\tilde{\omega}^2_3-3L^4\,\tilde{\omega}_1^2}
\pm L^2\,\tilde{\omega}_1} { \tilde{\omega}^2_3 -4L^4 \,
\tilde{\omega}_1^2}
\nonumber
\\
& &
\label{eq34}
\\
a^2_{\pm} & = & \frac{1}{4 b_{\pm}(m_1\pm b_{\pm}L^2\Gamma)} \,\, ,
\nonumber
\end{eqnarray}
where
\begin{equation}
\tilde{\omega}_3^2= \frac{\tilde{m}_3}{m_1} +\frac{L}{4} \omega_c^2
\label{eq34a}
\end{equation}
\begin{equation}
\tilde{\omega}_1 = \frac{\Gamma}{m_1} \equiv \frac{1}{m_1}
(L-1) \int d\vec{r} \, r^{2L-4} \gamma(r)
\label{eq35}
\end{equation}
\begin{equation}
m_1=  L^2  \int d \vec{r} \, r^{2L-2} \rho (r)
\label{eq36}
\end{equation}
\begin{equation}
\tilde{m}_3= \tilde{m}_3(T)+\tilde{m}_3(ee)+ \tilde{m}_3(+e)
\label{eq37}
\end{equation}
with
\begin{equation}
\tilde{m}_3(T)  =  L^2(L-1) \int d \vec{r} \, r^{2L-4}
 \left[L \tau + 2(L-2) \lambda \right]
\label{eq38}
\end{equation}
\begin{eqnarray}
\tilde{m}_3(ee) & = & 4 \pi L^2\frac{(2L-1)!!}{2^LL!}
\int_0^\infty  \rho'(r)dr \left\{\frac{1}{r}
\int^r_0 \left[2(L+1) \rho'(r') r'^{2L}+\rho''(r')r'^{2L+1}\right]
E_L\left(\frac{r'}{r}\right) dr' \right.
\nonumber
\\
& + & \left. r^{2L}\int^\infty_r
\left[\frac{\rho'(r')
}{r'}-\rho''(r')\right]E_L\left(\frac{r}{r'}\right)
dr'- \frac{2^{L+1} L!}{(2L+1)!!} r^{2L} \rho'(r)\right\}
\nonumber
\\
& - & 2 \pi L^2\int^\infty_0 dr\, \rho'(r) \left\{r^{2L-3}\int^r_0
\left[4  r'^2\rho'(r')+ r'^3 \rho''(r')\right]
E_1\left(\frac{r'}{r}\right)\,dr'\right.
\nonumber
\\
& + & \left. r^{2L}\int^\infty_r
\left[\frac{\rho'(r')}{r'}-\rho''(r')\right] E_1
\left(\frac{r}{r'}\right) \,dr'- \frac{4}{3} r^{2L} \rho'(r) \right\}
\label{eq39}
\end{eqnarray}
\begin{equation}
\tilde{m}_3(+e)=-\pi L^2
\int^\infty_0 dr V_+(r)\left[(2L-1)r^{2L-2}
\rho'(r)+ r^{2L-1} \rho''\right] \,\, .
\label{eq40}
\end{equation}
Besides  the particle and kinetic energy densities
\begin{equation}
\rho (r)=\sum_{nl\sigma} u^2_{nl\sigma}(r)
\nonumber
\end{equation}
\begin{equation}
\tau
(r)=\langle 0 \mid\sum_i \stackrel{\leftarrow }{\bigtriangledown}
\delta(\vec{r}-\vec{r_i})\stackrel{\rightarrow}{\bigtriangledown}
\mid 0 \rangle =
\sum_{nl\sigma}(u'^2_{nl\sigma}\,\,+\frac{l^2}{r^2}
u^2_{nl\sigma})  \,\, ,
\label{eq32}
\end{equation}
we have introduced in these equations the  densities
\begin{equation}
\gamma (r) = - \,\sum_{n\ell \sigma} \ell u^2_{n \ell \sigma}(r)
\, = \, r \,j_p(r)
\label{eq41}
\end{equation}
\begin{equation}
\lambda (r) = \sum_{n\ell \sigma} \frac{\ell^2}{r^2} u^2_{n \ell
\sigma}(r)
\label{eq43}
\end{equation}
as well as the E$_L$-function:
\begin{equation}
E_L(z)\equiv \frac{\pi}{2} F\left(-\frac{1}{2}, L+\frac{1}{2};L+1;z^2\right)
 \,\, ,
\label{eq44}
\end{equation}
where F is the hypergeometrical function \cite{gra}.
The density $\lambda$(r) represents
 a sort of centrifugal kinetic energy density. It can be
shown that in the zeroth-order Thomas-Fermi (TF) approximation
$\lambda$(r)=$\frac{1}{2} \tau$(r), similar to the
$\lambda$ (r)=$\frac{2}{3} \tau$(r) relation holding in three
dimensions \cite{bub}. The function j$_p$(r) is the paramagnetic
current density \cite{fer}. It is worth to see that at B=0,
$\gamma$(r) vanishes due to time reversal invariance.
Its contribution at high magnetic fields is crucial to have
well behaved B-dependent collective energies.
Eqs. (\ref{eq33}-\ref{eq40}) constitute the main outcome of the present
Section and will be used in Section VI to obtain numerical results
within CDFT.

The goal of describing the multipole modes in a situation as general
as possible makes these expressions to look very cumbersome at first
sight. However, it can be noticed that:

i) For L=1 and the parabolic V$_+$, one recovers the exact solution.

ii) For any L and zero magnetic field, b$_{\pm}=1/\tilde{\omega}_3$
and O$^+_{\pm L}\mid$0$\rangle$ merge into a degenerate state of energy
\begin{equation}
\omega_{\pm L}= \sqrt{\frac{\tilde{m}_3}{m_1}} \,\, .
\label{eq45}
\end{equation}
This is precisely the B=0 scaling energy E$_3$(Q$_L$), since
m$_1$ is actually  the general m$_1$(Q$_L$) sum rule, and
as it is outlined in the Appendix,
$\tilde{m}_3$ reduces to the zero magnetic field
m$_3$(Q$_L$) sum rule making $\gamma$(r)=0.
This result is the variational analog of the case we have
discussed for the dipole mode in a parabolic V$_+$.

iii) In the L=1 case, irrespective of B all terms in $\tilde{m}_3$
but $\tilde{m}_3(+e)$ vanish, and for  any axially symmetric
ionic potential V$_+$(r) we find
\begin{equation}
\omega_{\pm 1}= \sqrt{ \frac{\omega_c^2}{4} +
\frac{1}{2 N} \int d\vec{r} \,\Delta V_+(r) \,\rho(r)}
 \pm \frac{\omega_c}{2} \,\, ,
\label{eq45a}
\end{equation}
which reduces again to the exact case if V$_+=\frac{1}{2}
\omega_0^2$r$^2$.

iv) For any B, the states $\mid \pm L\rangle$ and their energies
satisfy the $m_1$ sum rule, i.e.:
\begin{equation}
m_1(Q_L)= L^2 \int \,d\vec{r}\, r^{2L-2} \rho(r)=
\omega_+ \mid \langle 0\mid [Q_L,O^+_{+L}]\mid 0\rangle\mid^2+
\,\omega_-\mid
\langle 0 \mid [Q_L,O^+_{-L}]\mid 0 \rangle \mid^2\,\,\, .
\label{eq46}
\end{equation}
It implies that, although O$^+_{\pm L}\mid$0$\rangle$ are only
approximate solutions to the L-polar excitation spectrum, within our
method there are no other excited collective states corresponding to
that multipolarity built as coherent superposition of one electron-hole
states.
The sum rule
\begin{equation}
m_3(Q_L)=\tilde{m_3}(T)+\frac{3}{4} L^2
\omega_c^2 m_1+3  \omega_c L^3
\Gamma+\tilde{m_3}(ee)+\tilde{m_3}(+e)
\label{eq47a}
\end{equation}
is also fulfilled. Moreover, the square transition
amplitudes to the $\mid\pm$L$\rangle$ states are equal,
\begin{equation}
\mid \langle 0\mid [Q_L,O^+_{+L}]\mid 0\rangle\mid^2 \,=  \,
\mid\langle 0\mid [Q_L,O^+_{-L}] \mid 0\rangle\mid^2 \,\, .
\label{eq47}
\end{equation}
The fulfillment of m$_1$ and m$_3$ constitutes a rather stringent test
on our variational method.

v) It is worth to notice that in all cases, there is no explicit
contribution to these formulas arising
from the exchange  correlation energy terms in the CDFT
Hamiltonian. This is because Q$_L$ is a solution of the
two-dimensional Laplace equation. A similar result
holds in three dimensions within LDA if
one takes for Q$_L$ a solution of the corresponding Laplace equation
\cite{ser4}.

\section{Edge waves in neutral large dots at B=0}

It is possible to use the scaling energy E$_3$, which
is a good approximation to the collective excitation energy at
B=0, to obtain the dispersion relation of the edge waves in
the case of a neutral large dot.

Under these conditions, the g.s. electronic density is
constant everywhere apart from a narrow region along
the border of the disk. Let $\rho_0$=N/($\pi$R$^2$) be that density, and
let $a$ be the width of the edge region. Using the zeroth-order TF
approximation to the kinetic and centrifugal kinetic energy densities
\begin{eqnarray}
\tau & = & \pi \rho^2
\nonumber
\\
& &
\label{eq48}
\\
\lambda & = & \frac{1}{2} \tau   \,\, ,
\nonumber
\end{eqnarray}
one gets from Eqs. (\ref{eq36},\ref{eq38}-\ref{eq40}):
\begin{equation}
m_1=    \pi \rho_0 L R^{2L}
\label{eq49}
\end{equation}
\begin{equation}
\tilde{m}_3(T)=2 L^2(L-1) \pi^2 \rho_0^2 R^{2L-2}
\label{eq50}
\end{equation}
\begin{equation}
\tilde{m}_3(ee) + \tilde{m}_3(+e)=
4\pi L^2 R^{2L-1}\, \rho^2_0\frac{(2L-1)!!}{2^L L!} F_L(z)  \,\, ,
\label{eq51}
\end{equation}
with  z=1$-{\em O}(a$/R) at least, and
\begin{equation}
F_L(z)\equiv \frac{\pi}{2}F\left(\frac{1}{2},L+\frac{1}{2};L+1;z^2\right)
\,\, ,
\label{eq52}
\end{equation}
which diverges at z=1. Physically,
this divergency is associated with the known divergency of the
electric field at the edge of the disk in the case of a constant
electron density \cite{gio}, see below.

Thus,
\begin{equation}
E^2_3=2 \pi \rho_0 \frac{L(L-1)}{R^2}+
4 \frac{\rho_0}{R} \frac{(2L-1)!!}{2^L(L-1)!} F_L(z) \,\, .
\label{eq53}
\end{equation}
F$_L$(z)  can be written as a function of F$_0$(z), which is the
elliptic function $\bf{K}$(z) \cite{gra}.
\begin{equation}
\frac{(2L-1)!!}{2^LL!} F_L(z) = F_0(z)
- \sum^L_{m=1} \frac{1}{2m-1} \,\, .
\label{eq54}
\end{equation}
Using it we get
\begin{equation}
E^2_3=2\pi \rho_0\frac{L(L-1)}{R^2} +4 \rho_0\frac{L}{R}
\left[\frac{1}{2} F_1(z)+1- \sum^L_{m=1}\frac{1}{2m-1}\right] \,\,.
\label{eq57}
\end{equation}
On the other hand, F$_1$(z) can be related to the
electric field E(r) generated by the electrons at the edge of the disk:
\begin{equation}
2\rho_0 F_1(z)= -\frac{dV}{dr}\mid_{r \rightarrow
R} = E(r)_{r\rightarrow R} \,\, .
\label{eq58}
\end{equation}
The divergency of E(r) at r=R is removed when one considers the
existence of the edge widthness \cite{gio}:
\begin{equation}
E(R)= 2 \rho_0 \,\, ln\,\left( \gamma \frac{R}{a}\right)   \,\, ,
\label{eq59}
\end{equation}
where $\gamma$ is a constant that depends on the precise way the
electronic density goes to zero at R. Thus,
\begin{equation}
E^2_3=2 \pi \rho_0 \frac{L(L-1)}{R^2}+ 4 \rho_0\frac{L}{R}\,\left[
\frac{1}{2}\,\, ln\,\left(\gamma \frac{R}{a}\right)
+1-\sum^L_{m=1}\frac{ 1}{2m-1} \right]  \,\, .
\label{eq60}
\end{equation}
If L$>>$1 but still N$>>$L, that expression can be further elaborated
since
\begin{equation}
\sum^L_{m=1}\frac{1}{2m-1} \sim \frac{1}{2}(C+\,ln\,L)\,+\,ln\,2\,\,\, ,
\label{eq61}
\end{equation}
where C is the Euler constant 0.5772...
Introducing the wave vector q = L/R, and keeping only the
leading q-term we get
\begin{equation}
E_{3} =   \omega(q)= \sqrt{2  \rho_0 \,q\,\ln\,(q_0/q)} \,\, ,
\label{eq62}
\end{equation}
where $q_0=\gamma/(a \beta)$ and $\beta = 0.964$.

Result Eq. (\ref{eq62}) can also be obtained using classical
hydrodynamics
\cite{gio}. It is worthwhile to notice that the above $\omega$(q)
 differs from that obtained within the edge-magnetoplasmon
model (Refs. \cite{gla,mar,san}):
\begin{equation}
\omega(q)= \sqrt{\frac{4 \pi}{3} \,\rho_0 \, q}\,\, ,
\label{eq63}
\end{equation}
which is $\sqrt{2/3}$ times that of the two-dimensional plasma
frequency. We would
like to stress that to get the $\sqrt{q \ln (q_0/q)}$ dispersion
relation, it is crucial to take into account widthness effects in
the electronic density. These effects are important in quantum dots,
where the number of electrons usually is much less than the number of
ions, and in any case the electron density has a non negligible edge
region.

\section{Numerical results}

We have applied the method of Section IV to dots made of N= 6, 20,
30, 42, and 56 electrons.
We have taken the values of g$^*, \epsilon$ and m$^*$ indicated after
Eq. (\ref{eq1}), and N$_+$=125, i.e., a positive density
of $\sim$ 4 $\cdot$ 10$^{11} $ions/cm$^2$. This corresponds to a GaAs
disk of about 1000 $\AA$ radius modelling the positive background.

The g.s. of the dots has been obtained using the CDFT of Ref \cite{fer}.
We have checked that we reproduce their results when we use the same
V$_+$. For large dots and intense magnetic fields, sometimes
one has to face  severe convergence
problems in the solution of the Kohn-Sham equations.
Rather than a  deficiency of the numerical algorithm, we
consider it as a consequence of a inherent
characteristic of the system under study, namely the existence of
a very dense s.p. energy spectrum.
To overcome it on one hand, and to carry out the calculations under
conditions closer to the experimental ones on the other hand,
we have found it convenient to approach the description of the
collective spectrum as the low-temperature limit of the results
obtained from the finite temperature generalization of the formalism
of Section II (see Ref. \cite{bar}), and of the Kohn-Sham equations
(see for example, Ref. \cite{fer2}). Consequently,
the numerical results we discuss below have been obtained  at a
temperature T $\sim$ 1-2 K. A comparison with several cases in which
the T=0 calculation is easy to converge, allows us to state that the
small temperature we use does not influence the results here presented.
Thermal effects on the collective spectrum of quantum dots will
be described in details elsewhere.

Figures \ref{fig1}-\ref{fig2} show the electronic densities
corresponding to dots with N=6, 30 and 56 electrons for B=0, and
for B=5 tesla, respectively.
Figures \ref{fig3}-\ref{fig5} display the B-dependent, multipolar 
spectrum of the same dots up to L=4. This is the interesting region 
where the crossing between  $\omega_{+L}$ and $\omega_{-L'}$ branches
may occur and has been experimentally observed. The energies are
drawn in meV, and the magnetic field in tesla. These figures
show that as the $\omega_{-L}$ energies go to zero, they may reach
a value comparable to the electron-hole s.p. energy difference at a
rather moderate B-value. When this happens, the collective state
lies within the s.p. excitations region  and experiences
a strong Landau damping, loosing its collectivity and eventually being
washed out. This is one of the reasons why the $\omega_{+L}$
branch has been experimentally observed up to higher B-values
than the $\omega_{-L}$ branch \cite{dem}.

The structures appearing along the L$>$1 curves are due to drastic
changes in the particle, kinetic and paramagnetic current
densities of the dot arising from the effect of B
through the exchage-correlation energy, which has a profound
influence on them (see Fig. 3 of Ref. \cite{fer}). The structures in
the $\omega_{\pm L}$ branches roughly correspond to values of B at which
the total g.s. spin has a minimum. For example, for N=6 we have that 
2S is equal to 2 at
B=1 T, to zero at B=2 T, and to 2 at B$\sim$3 T. For N=56, one has
that 2S=13 at B=4 T, 7 at B$\sim$5 T, and 8 at B=6 T. For the N=6
dot, the rising of the $\omega_{+L>1}$ curves
at B$\sim$5 T is due to the
full alignment of the electron spins. No similar risings show up
for N=30 and 56 because for them, the alignment occurs at B-values
higher than those displayed in the figures.

It can be seen from Figs. \ref{fig3}-\ref{fig5}
that the crossing between  $\omega_{+1}$ and $\omega_{-L}$
branches does not follow a clear N-systematics.
We have also plotted in Fig. \ref{fig4} the  E$_3$(Q$_{L}$)
energies for L=1 and 4 (dashed lines). The scaling
energy is reproducing $\omega_{+L}$ to within 10-20 \%, since the
$\omega_{-L}$ branch is contributing to m$_3$(Q$_L$) with the same
weight as the $\omega_{+L}$ one, and the negative B-dispersion
energies are going to zero rather slowly.

We also display
in Fig. \ref{fig5} (dashed lines), the collective energies obtained
for a parabolic potential whose $\omega_0$ has been fixed to
5.6 meV in order to reproduce the $\omega_1$ energy at B=0.
It is worth to notice that this value does not equal  the one
which fits the Coulomb potential generated by the 1000 $\AA$
radius disk
charged with N$_+$=125 ions near the origin, which is 4.4 meV, and
consequently, $\omega_0$ has to be interpreted here as an effective
parameter to reproduce the dipole energy at zero magnetic field.
For L=1, to the scale of the figure both calculations coincide.
It may be seen that the B-slopes of the
$\omega_{\pm L}$ branches for L=2 to 4 are roughly the same for
the parabolic and disk confining potentials.

At B=0, Figs. \ref{fig6}-\ref{fig7} show for N=6 and 56
respectively, the different relative
contributions to m$_3$ coming from  kinetic and  Coulomb energies,
as a function of L. The Coulomb energy
is also decomposed into  e-e and dot-electron (+e) energies.
These figures show that for small L-values, m$_3$ is dominated by the
(+e) contribution, the kinetic and (e-e) ones being of minor
importance. However, for a fixed N the kinetic  contribution is
eventually taking over the Coulomb energy contribution. It occurs at
an angular momentum L$_{cr}$ which increases when N increases.
Since  Q$_L$  behaves as
$\sim$ r$^L$, the higher the L, the more external its influence on the
electronic density. It means that for large enough L, it
just acts on the s.p. wave functions having the bigger l-angular
momentum and radial quantum number n. Consequently, it no longer
generates collective but s.p. excitations.
As collectivity has its origin in interparticle effects,
it has been argued in Ref. \cite{ser} that L$_{cr}$ roughly represents
the largest L collective mode the system can sustain, since for
L $>$ L$_{cr}$
the restoring force represented by m$_3$ is basically determined
by an independent particle property like the kinetic energy. This
criterion yields L$_{cr}\sim$ 4, 6 and 8 for N=6, 30 and 56,
respectively.

We represent in Fig. \ref{fig8}  the
energy of modes with L=1 to 4 at B=0, as a function of the
number of electrons in the dot. The ratios $\omega_3/\omega_1$
and $\omega_2/\omega_1$  have, for N$>$20, average values 1.68 and 1.37
respectively, instead of 3 and 2 as corresponds
to the harmonic oscillator sequence. It is apparent from that figure
the N-dependence of $\omega_L$. As we have already discussed,
only for the dipole mode in the case of a parabolic confining potential
one has $\omega_1 = \omega_0$. The
deviation from that rule becomes more and more important as the number
of electrons in the dot increases. This has been already discussed in
Ref. \cite{bro}. The N-dependence of $\omega_{L>1}$ is more complex,
and it is due to the interplay between kinetic and Coulomb energy
contributions to the excitation energy.
It can be easily understood from Eq. (\ref{eq53}), which shows
that the contribution of the kinetic energy term
is important for small N and increases with increasing L. For
fixed L, the Coulomb energy eventually dominates, and $\omega_{L>1}$
depends on N in a way similar to the dipole mode.

Finally, we have also studied the collective spectrum of
the N=6 dot in the parabolic potential with $\omega_0$=5.6 meV,
and have found that as
expected, the $\omega_{L>1}$ energies at B=0 depend on N. For example,
$\omega_2$ decreases from 8.2 to 7.5 meV, $\omega_3$ decreases from
10.9 to 9.0 meV, and $\omega_4$ decreases from 14.3 to 10.4 meV when
one goes from N=6 to N=56.

So far, we have presented a  systematic study carried out under
well defined conditions which could render it difficult the comparison
with the results of a given experiment, since there are several
variables that have to be fixed at the  experimental values to
permit a sensible comparison. To end this section, we want
to compare some results obtained within our formalism
with the  experimental data of Ref. \cite{dem}. This is possible
only in part because in that work, the L=1
and 2 well defined branches have been obtained for a
dot made of a large number of electrons, N=210. At present,
this is too large a value for us to deal with microscopically.
Let us remind that the s.p. wave functions behave near the origin
as r$^{\mid l \mid}$ (see for example Ref. \cite{mak}),
and since the s.p. levels are nondegenerate
when B$\neq$0,  large N's imply large s.p. angular momenta.
Consequently, we have only attempted to describe the N=25, R=1000
$\AA$ dot.
It can be inferred from the value of the dipole energy
Demel et al. find at B=0, that N$_+ \sim $ 28.
 For this dot, the ratio $\omega_2/\omega_1$ they obtain
is around 2, a value we are unable to reproduce, whereas
for the N=210 dot it is $\sim$1.5, in better agreement with our
systematics. Nevertheless, it can be seen from Fig. \ref{fig9} that
$\omega_{+1}$ and E$_3$(Q$_1$)  nicely reproduce the positive
B-dispersion branch, which is the only one for which a detailed
comparison is possible.

\section{Summary}

In this work we have used a variational approach similar to the one
proposed by Feynmann in the case of liquid $^4$He, to describe the
multipole spectrum of quantum dots. One of the merits of the method is
that quantum and finite size effects can be taken into account.
It may be easily applied to dots hosting several tens of electrons.
Rather than in the method itself, this
limitation has its origin in technical difficulties inherent to
current microscopic approaches to handle a large number of electrons
in intense magnetic fields.

We have presented a systematic description of edge modes up to L=4
in the region of interest to describe  level  crossing at non zero
magnetic fields. We have given explicit formulas
for the m$_1$ and m$_3$ sum rules corresponding to
the general multipole operator Eq. (\ref{eq27}), which reduce to
very simple expressions in the dipole case, Eqs. (\ref{a24}).
These sum rules are exact within CDFT, and may be of interest to
check the accuracy of any detailed calculation of the L-mode strength,
in a similar way as they are currently used within TDLDA \cite{gue}.
Besides this practical application, it is worth to notice that
there are few studies in the literature of an m$_3$ sum rule
corresponding to a physical situation where time-reversal invariance
is violated (see for example Refs. \cite{god,raj} for the three
dimensional polarized electron gas).

For large neutral dots at zero magnetic field, we have shown that
the classical hydrodynamic dispersion law for edge waves
$\omega$(q) $\sim \, \sqrt{q\, \ln \,(q_0/q)}$  holds when
quantum and finite size effects are taken into account.
Finally, we have also shown that in the case of a parabolic potential,
the dipole mode can be exactly solved, yielding the well known classical
formula for $\omega_{\pm 1}$.
The exactness of the dipole collective spectrum was stressed in Refs.
\cite{mak,bak}. Here, we have obtained it in a different way, and
 have gone a step further expliciting the structure of the
$\mid\pm$1$\rangle$ collective states. This has allowed us to get a
variational solution valid for any axially symmetric lateral confining
potential.

\acknowledgements

It is a pleasure to thank Luis Brey for an useful correspondence.
This work has been performed under grants PB95-1249 and PB95-0492
from CICYT and SAB95-0388 from DGID, Spain, and GRQ94-1022 from
Generalitat of Catalunya.

\eject
\appendix
\section*{}

In this Appendix we give some hints about how to derive results Eqs.
(\ref{eq33}-\ref{eq40}), and the  sum rules
m$_1$ and m$_3$ corresponding to
 the external one-body operator Q$_L$ of
Eq. (\ref{eq27}), which are given in Eqs. (\ref{eq36}) and
(\ref{eq47a}). The details of the method we use here can be found in
Refs. \cite{lip,boh,ser,ser4}.

We fix our attention on the operator O$^+_{+L}$ of Eq. (\ref{eq29});
the operator O$^+_{-L}$ is handled similarly.
The double conmutator in the numerator of Eq. (\ref{eq7}) can be
decomposed in three pieces:
\begin{eqnarray}
[O_{+L},[H,O^+_{+L}]] & = & a^2_+
\left\{[Q^+_L,[H,Q_L]]+ib_+L\left([R^+_L,[H,Q_L]-[Q^+_L,[H,R_L]
\right)\right.
\nonumber
\\
& + & \left. b_+^2 L^2[R^+_L, [H,R_L]]\right\}
\label{a1}
\end{eqnarray}
Splitting the Hamiltonian Eq. (\ref{eq3}) into a one body term H$_0$
and a two-body term V=$\sum_{i<j}$ V ($\mid
\vec{r}_i-\vec{r}_j
\mid$), and using that Q$_L$ is a local operator which
commutates with V one gets:
\begin{equation}
[H,Q_L]=[H_0,Q_L]=-iLR_L+\frac{1}{2} L \omega_cQ_L \,\, ,
\label{a2}
\end{equation}
where we have used the conmutation relations
\begin{equation}
[T,Q_L]=-iLR_L\,\,,\,\,\,\,\,\,\,\,\,\,[L_z,Q_L]=LQ_L\,\,\,\,\,.
\label{a3}
\end{equation}
From Eq. (\ref{a2}) and
\begin{equation}
[Q_L^+,P]=2iLQ^+_{L-1}\,\,\,\,\,  ,  \,\,\,\,\,[L_z,P]=P\,\,\,,
\label{a4}
\end{equation}
it is then inmediate to derive for the first two terms of
Eq. (\ref{a1}) the following results:
\begin{equation}
\langle 0 \mid[Q_L^+,[H,Q_L]] \mid 0 \rangle = 2 L^2 \int r^{2L-2}
\rho (r) \,d \vec{r}\,\, =\,2 m_1  \,\,
\label{a5}
\end{equation}
\begin{equation}
\langle 0 \mid[R_L^+,[H,Q_L]]-[Q^+_L,[H,R_L]] \mid 0 \rangle = -2i
(\omega_c m_1+4L\Gamma)   \,\, ,
\label{a6}
\end{equation}
where
\begin{equation}
\Gamma= \frac{1}{4} \langle 0 \mid [R^+_L,R_L] \mid 0 \rangle = (L-1)
\int d\vec{r}\, r^{2L-4}\, \gamma (r) \,\, .
\label{a7}
\end{equation}
To evaluate  $[R^+_L,[H,R_L]]$ we first calculate
\begin{equation}
[H,R_L]=[H_0,R_L]+[V,R_L]
\label{a9}
\end{equation}
The first term gives
\begin{equation}
[H_0,R_L]=-i(L-1)\sum^N_{i=1} q_{L-2}p^2 + \frac{1}{2}
L\omega_cR_L+\frac{i}{4}
\omega^2_cQ_L+i\sum^N_{i=1}(\partial_x+i\partial_y) V_+(r_i)q_{L-1}
\,\,
,
\label{a10}
\end{equation}
and it is then straightforward to recover the results of Eqs.
(\ref{eq38}) and (\ref{eq40}) for
\begin{equation}
\tilde{m}_3(T)=\frac{L^2}{2} \langle 0\mid [R^+_L,-i(L-1)\sum^N_ {i=1}
q_{L-2} p^2]\mid 0 \rangle
\label{a11}
\end{equation}
and
\begin{equation}
\tilde{m}_3(+e)=\frac{L^2}{2} \langle 0\mid [R^+_L\, , \,
i\sum^N_{i=1}
(\partial_x+i\partial_y) V_+(r_i) q_{L-1} ] \mid 0 \rangle \,\, .
\label{a12}
\end{equation}
The remaining two terms in Eq. (\ref{a10}) give:
\begin{equation}
\langle 0\mid [R^+_L \, , \, \frac{L}{2} \omega_cR_L]\mid
0\rangle\,=\,2L\omega_c\Gamma
\,\, ,
\label{a13}
\end{equation}
which yields the  Eq. (\ref{eq35}) term, and
\begin{equation}
\langle 0\mid [R^+_L,\frac{i}{4} \omega_c^2Q_L]\mid
0\rangle\,=\,\frac{1}{2}\frac{\omega_c^2}{L}m_1
\label{a14}
\end{equation}
which yields the second  term in Eq. (\ref{eq34a}).

It remains the problem of evaluating the  $\langle 0\mid
[R^+_L,[V,R_L]]\mid 0 \rangle$ term. We have done it
within CDFT  by scaling the CDFT g.s. as
\begin{equation}
\mid \eta \rangle =\,e^{\eta R_L} \mid 0 \rangle
\label{a15}
\end{equation}
and then calculating the derivative:
\begin{equation}
\frac{1}{2}\frac{\partial^2}{\partial \eta^2} \langle \eta \mid V
\mid \eta \rangle \mid_{\eta=0}\,\,\,\,,
\label{a16}
\end{equation}
where
\begin{equation}
\langle \eta \mid V\mid \eta \rangle = \frac{1}{2} \int
\frac{\rho_{\eta}(\vec{r}_1)\rho_{\eta}(\vec{r}_2)}{\mid
\vec{r}_1-\vec{r}_2\mid}
\,d\vec{r}_1\,d\vec{r}_2
\label{a17}
\end{equation}
and
\begin{equation}
\rho_{\eta}(\vec{r})=\langle \eta \mid \sum^N_{i=1}
\delta(\vec{r}-\vec{r}_i)\mid \eta \rangle\,=\,\rho\,+\,\eta
\rho_1\,+\,+\eta^2\rho_2\,+\,...
\label{a18}
\end{equation}
with
\begin{equation}
\rho_1=-Lr^{L-1} \rho'(r)\,e^{iL\theta}
\label{a19}
\end{equation}
\begin{equation}
\rho_2=\frac{1}{2} L^2(2L-1)r^{2L-3} \rho'(r)\,+\,
\frac{1}{2} L^2 r^{2L-2}\rho''  \,\,\, .
\label{a20}
\end{equation}
From Eqs. (\ref{a15}-\ref{a20}) one gets the result
Eq. (\ref{eq39}):
\begin{equation}
\tilde{m}_3(ee)\,=\,\frac{L^2}{2} \langle 0 \mid [R^+_L,[V,R_L]]\mid
0\rangle\,\, .
\label{a21}
\end{equation}
It is seen that the exchange correlation energy does not
give any explicit contribution to $\tilde{m}_3(ee)$. However, it
affects  g.s. magnitudes like $\rho$(r) and the other
densities, thus implicitly influencing all these quantities.

The conmutator $[O^+_{+L},O_{+L}]$ of the denominator of
Eq. (\ref{eq7}) is easily evaluated to be
\begin{equation}
\langle 0 \mid [O^+_{+L},O_{+L}]\mid 0 \rangle\,=\, a^2_+ \langle 0
\mid [Q_L-i b_+ L R_L\, , \,Q^+_L+i b_+ L R^+_L]\mid 0 \rangle\,=\, 4
a^2_+(b_+m_1+b_+^2L^2\Gamma)
\label{a22}
\end{equation}
The cubic energy weighted sum rule Eq. (\ref{eq10}) for the external
operator Q$_L$ is given by
\begin{equation}
m_3\,=\, \frac{1}{2}\langle 0\mid [[H,-i L R^+_L +\frac{1}{2}
L\omega_c Q^+_L],-i L R_L +\frac{1}{2} L \omega_c Q_L]\mid 0 \rangle
\label{a23}   \,\, ,
\end{equation}
where we have employed Eq. (\ref{a2}). Using the previous result it is
easy to recover the expression Eq. (\ref{eq47a}), which at B=0
reduces to $\tilde{m}_3$
Eq. (\ref{eq37}) with $\gamma$(r)=0.

Let us finally indicate that
for the dipole operator, the m$_1$ and m$_3$
sum rules for any value of B and an axially symmetric confining
potential V$_+$   have  the simple expressions:
\begin{eqnarray}
m_1(D) & = & \frac{N}{2}
\nonumber
\\
& &
\label{a24}
\\
m_3(D) & = & \frac{N}{2} \omega_c^2 + \frac{1}{4}
\int \Delta V_+(r) \rho(r) \, d\vec{r}  \,\,  .
\nonumber
\end{eqnarray}

\eject

\begin{figure}
\caption{ Electronic densities in (a$^*_0$)$^{-2}$ as a function of r
in a$^*_0$, for  dots with N=6, 30 and 56 electrons  and B=0.}
\label{fig1}
\end{figure}
\begin{figure}
\caption{Same as Fig. 1 for B= 5 tesla. }
\label{fig2}
\end{figure}
\begin{figure}
\caption{$\omega_{\pm L}$ energies in meV as a function of B in tesla
for L=1 to 4, corresponding to N=6. }
\label{fig3}
\end{figure}
\begin{figure}
\caption{Same as Fig. 3 for N=30. The dashed lines are the E$_3$(Q$_L$)
energies for L=1 and 4.}
\label{fig4}
\end{figure}
\begin{figure}
\caption{Same as Fig. 3 for N=56. The dashed lines represent
$\omega_{\pm L>1}$ obtained using a parabolic potential instead of the
one generated by the disk.}
\label{fig5}
\end{figure}
\begin{figure}
\caption{Decomposition of m$_3$(Q$_L$) into kinetic (solid line),
total Coulomb (dashed line), (e-e)-component (dash-dotted line), and
(+e)-component (dotted line) contributions
as a function of L for the  N=6 dot at B=0. The lines are drawn
to guide the eye.}
\label{fig6}
\end{figure}
\begin{figure}
\caption{Same as Fig. 6 for N=56. }
\label{fig7}
\end{figure}
\begin{figure}
\caption{Zero magnetic field $\omega_L$ energies
in meV for L=1 to 4, as a function of N.}
\label{fig8}
\end{figure}
\begin{figure}
\caption{$\omega_{\pm 1}$ (solid lines) and E$_3$(Q$_1$) (dashed line)
energies in meV as a function of B in tesla, for a dot of R=1000
$\AA$, N$_+$=28, and N=25. The values of $\epsilon$, g$^*$ and
m$^*$ are those of GaAs given in the text. The points are
experimental results taken from Ref. [6].}
\label{fig9}
\end{figure}
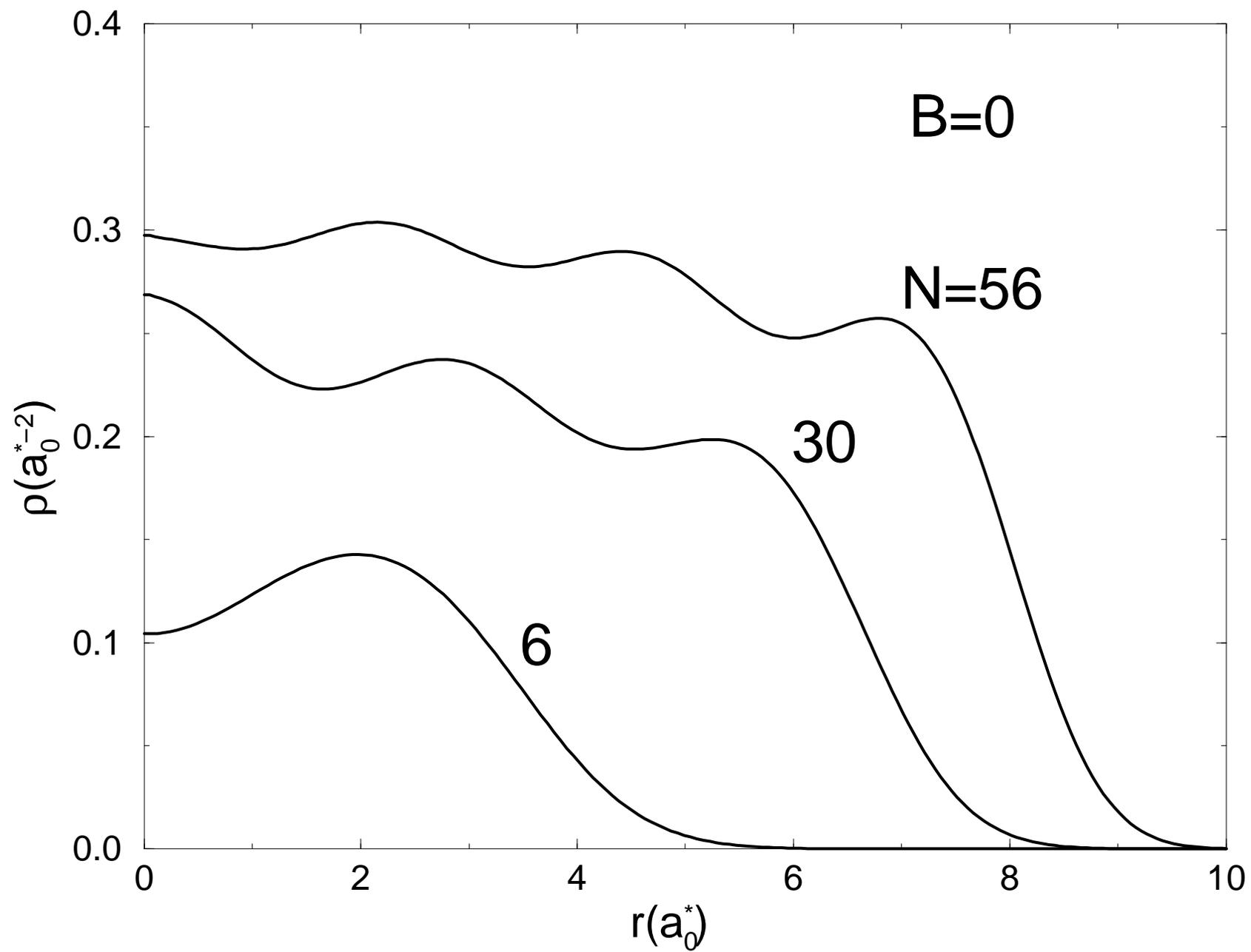
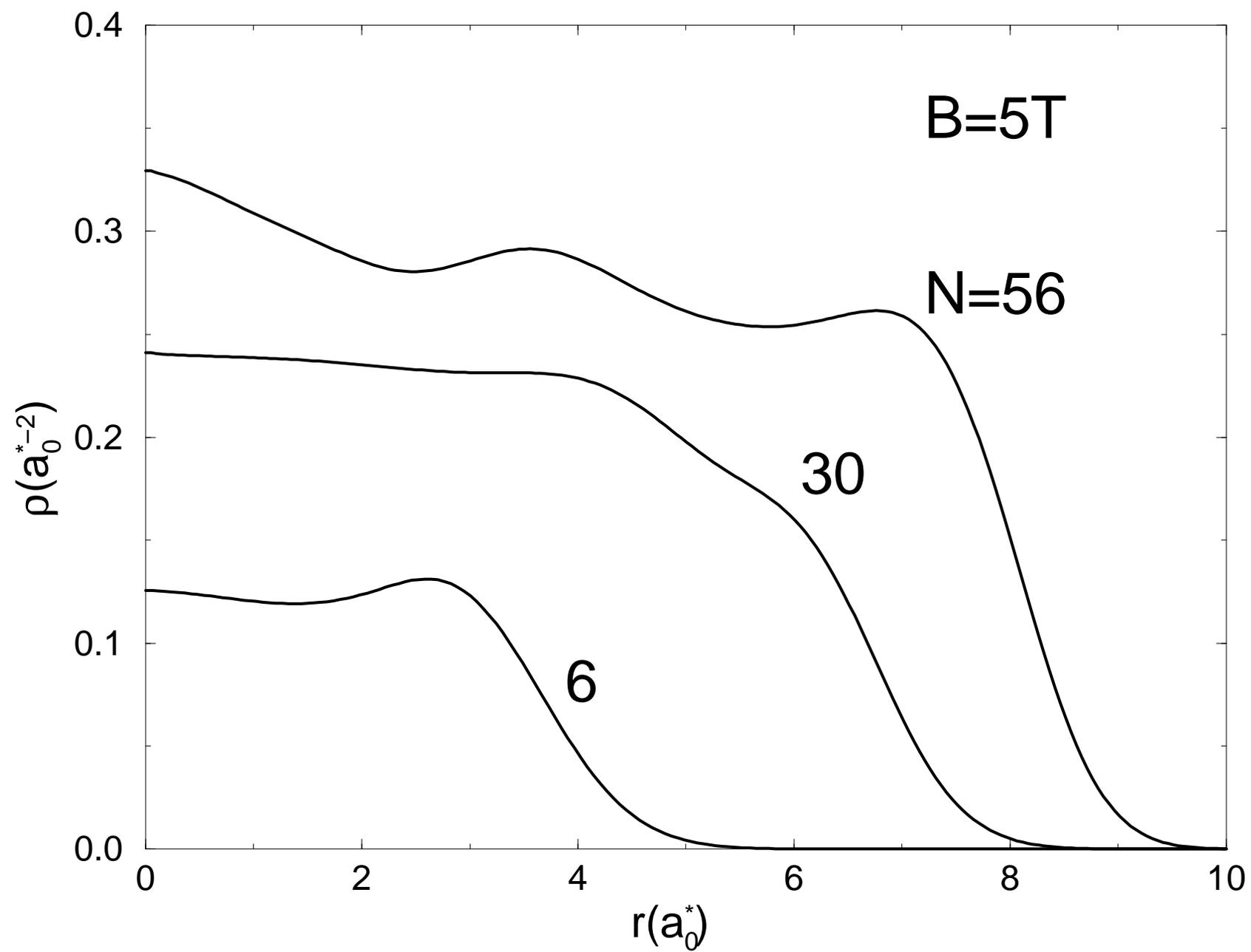
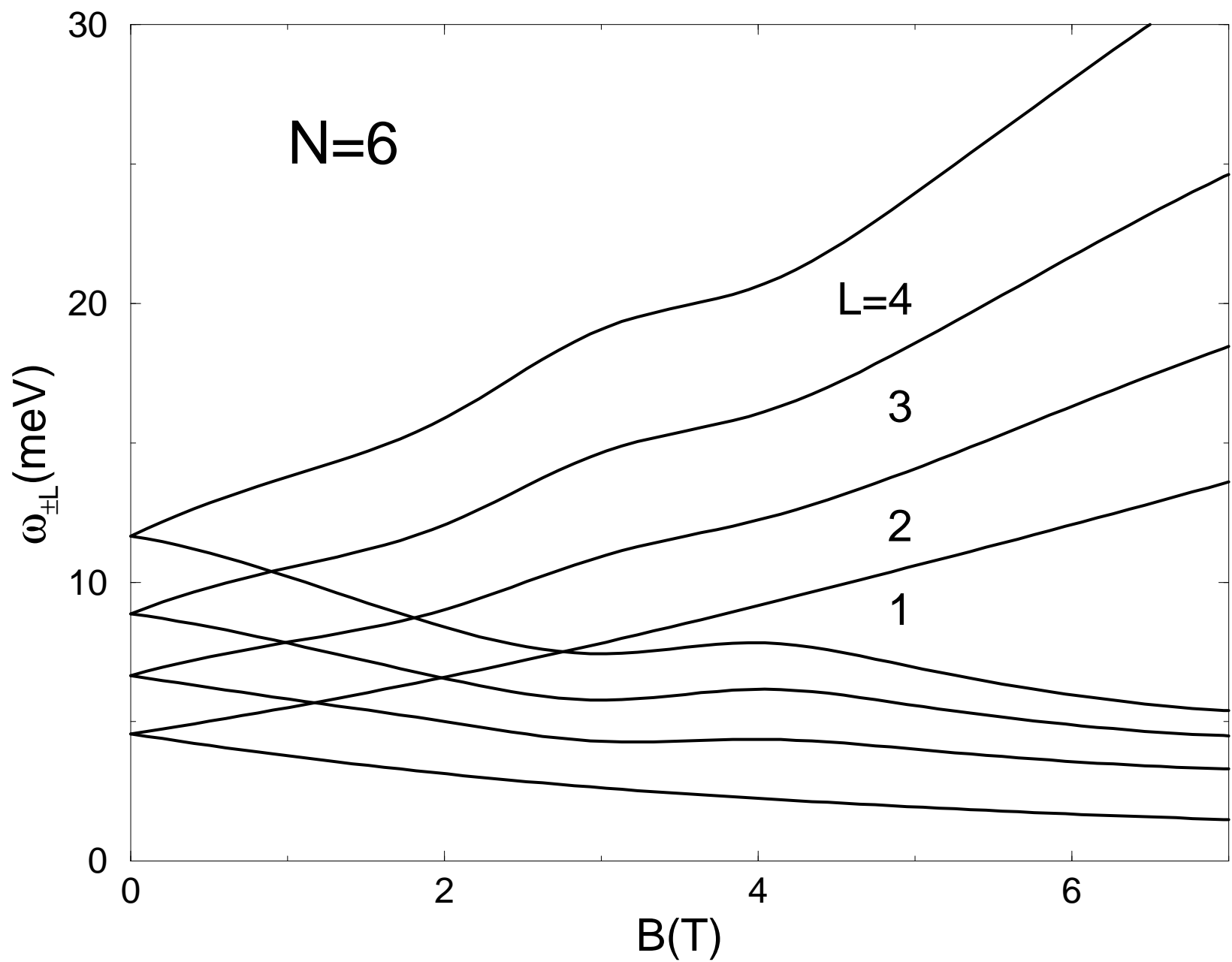
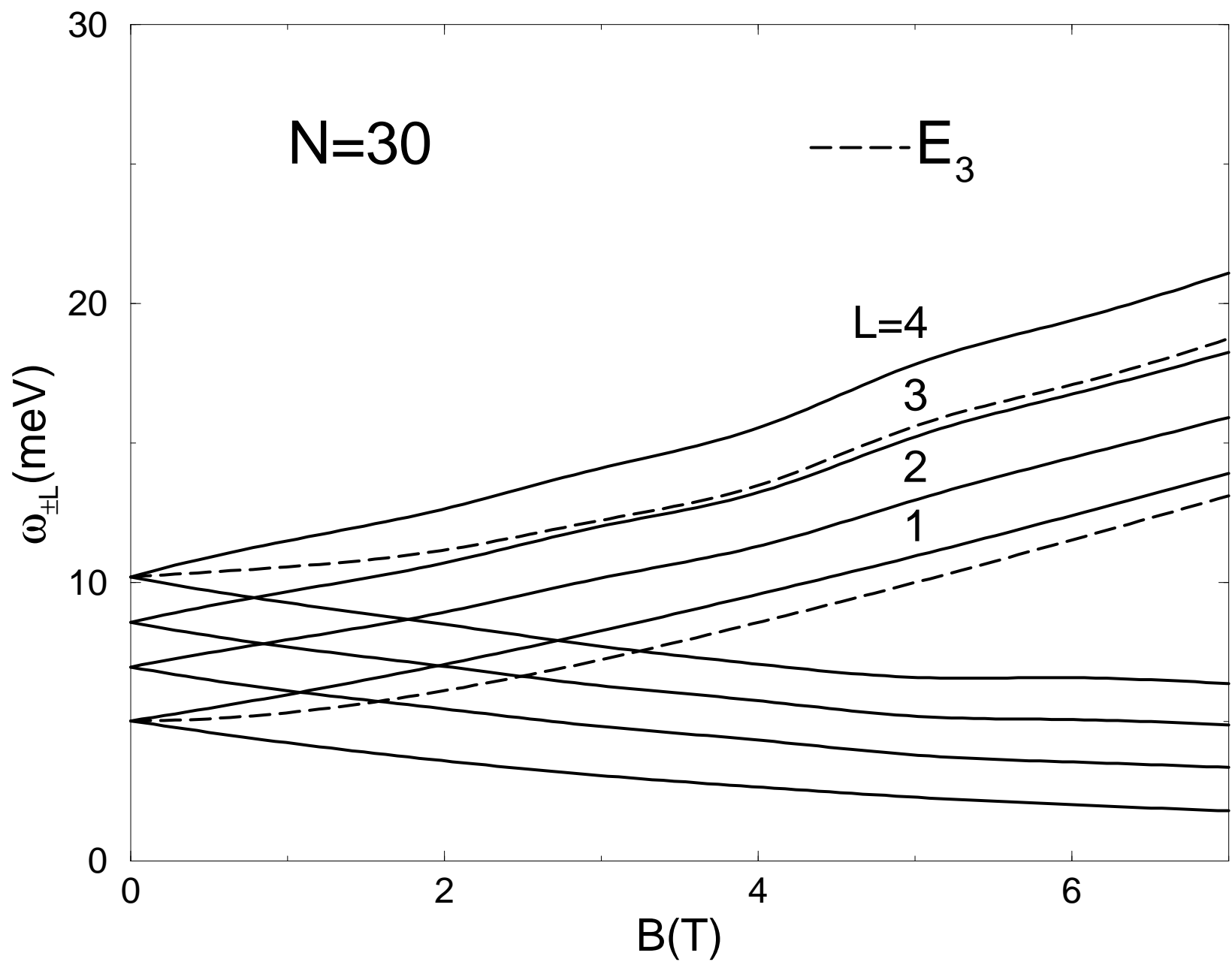
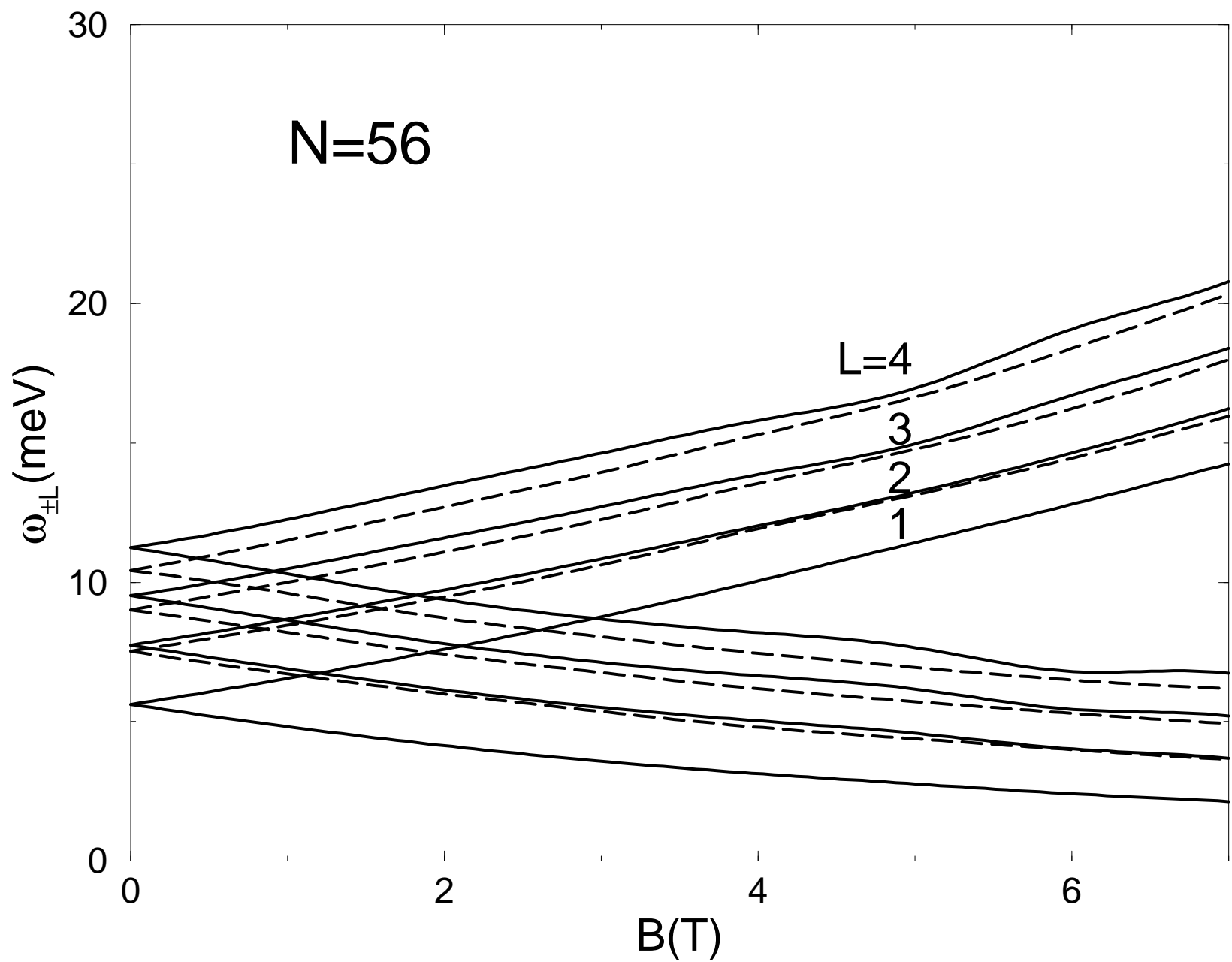
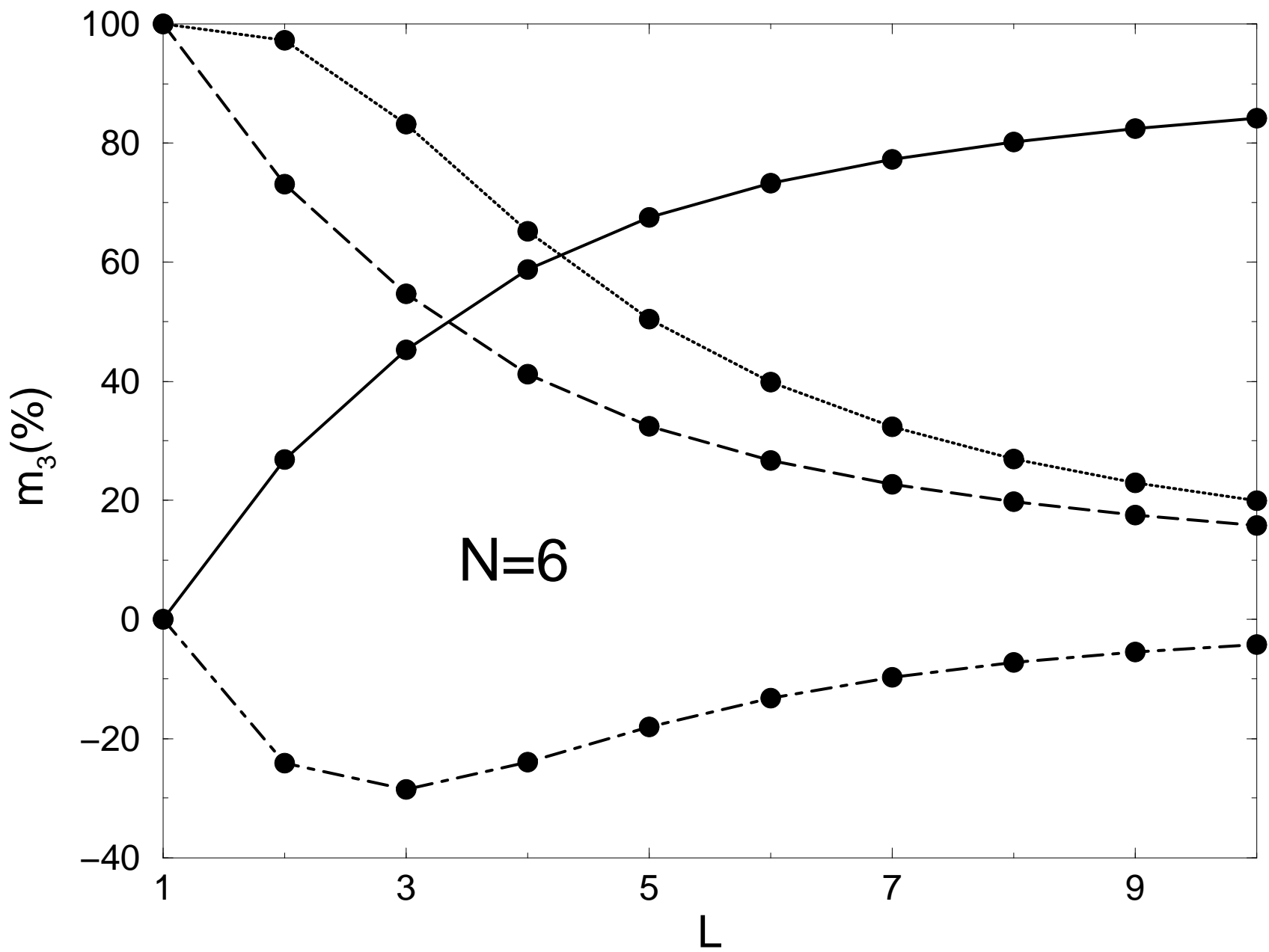
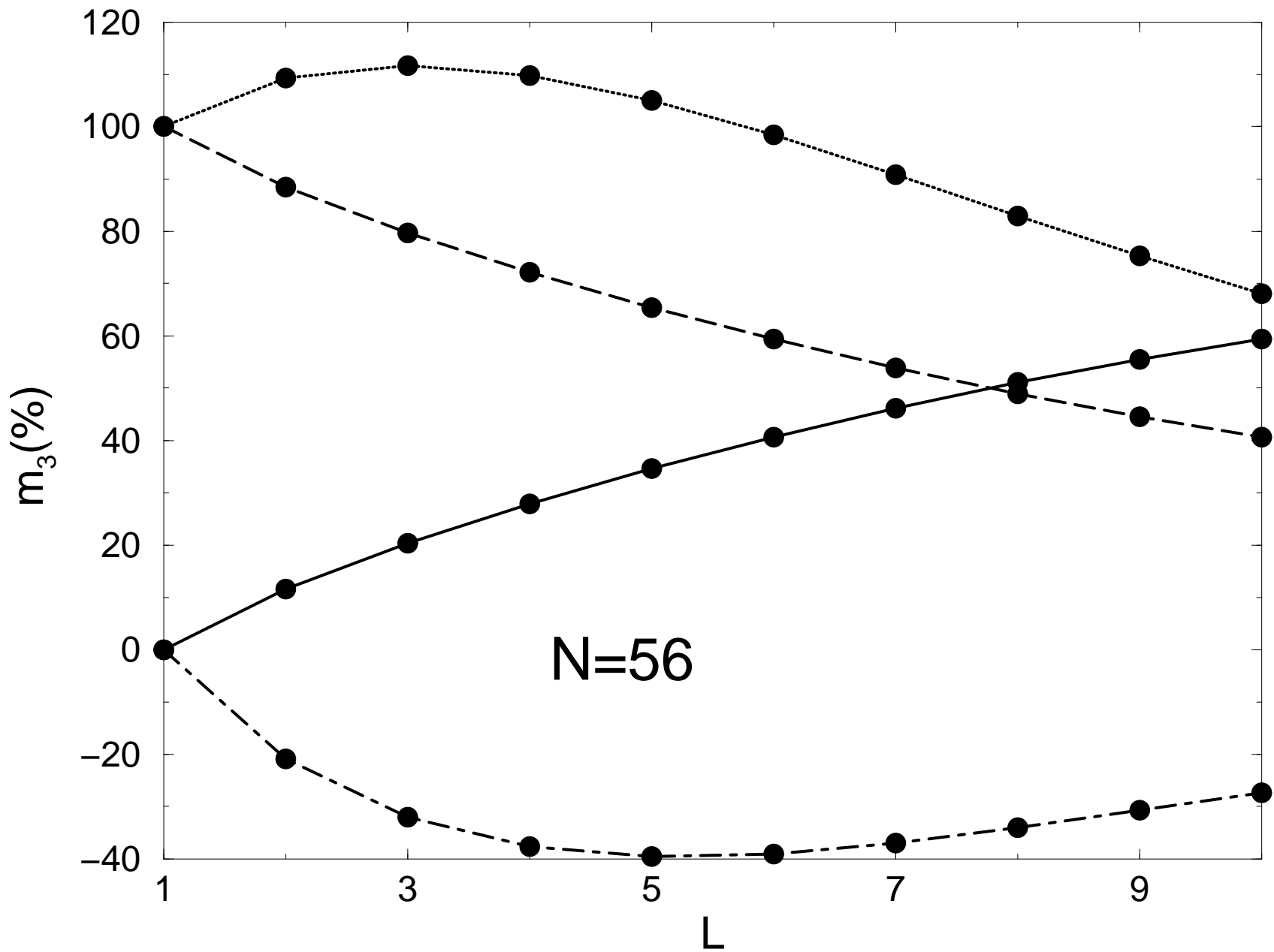
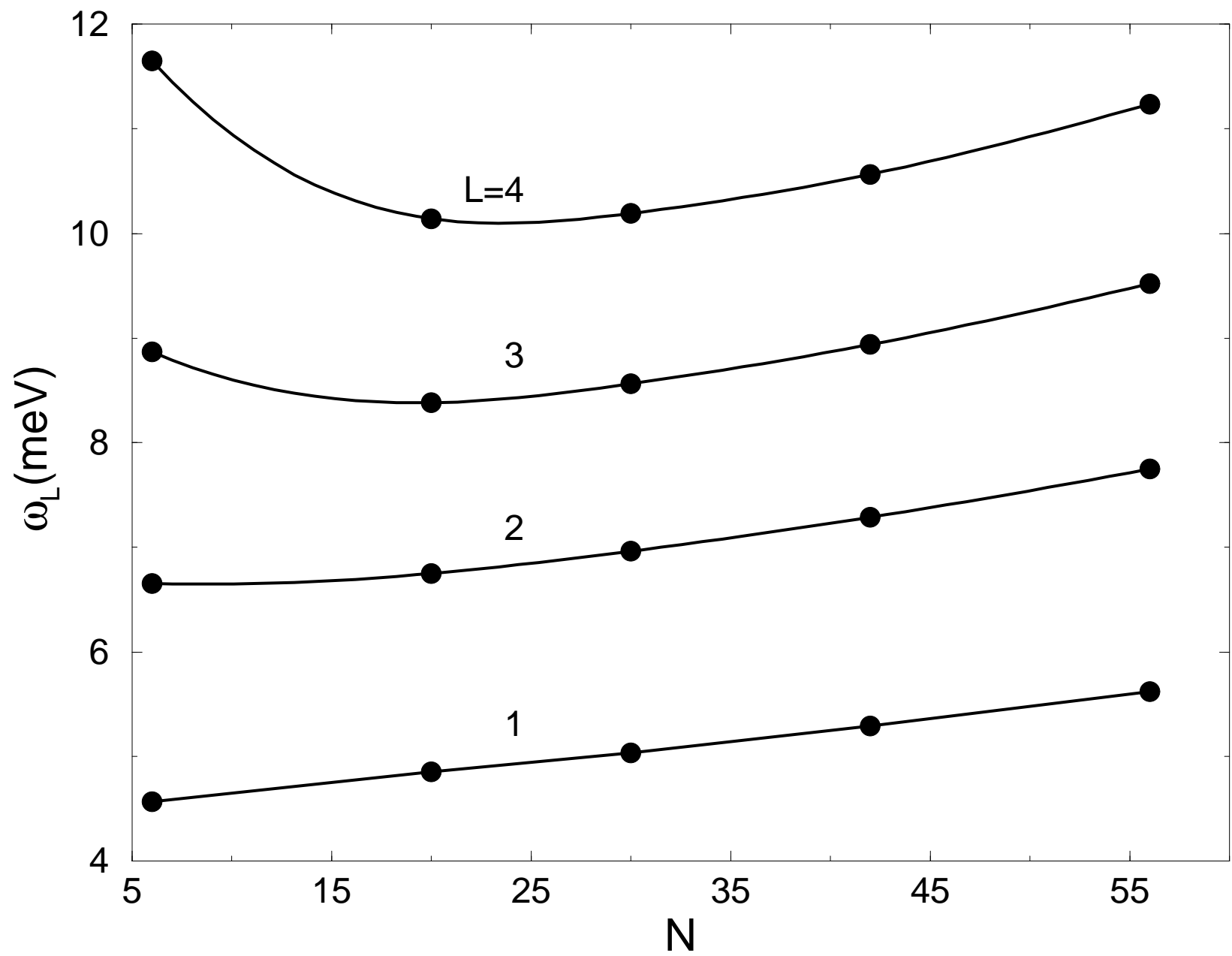
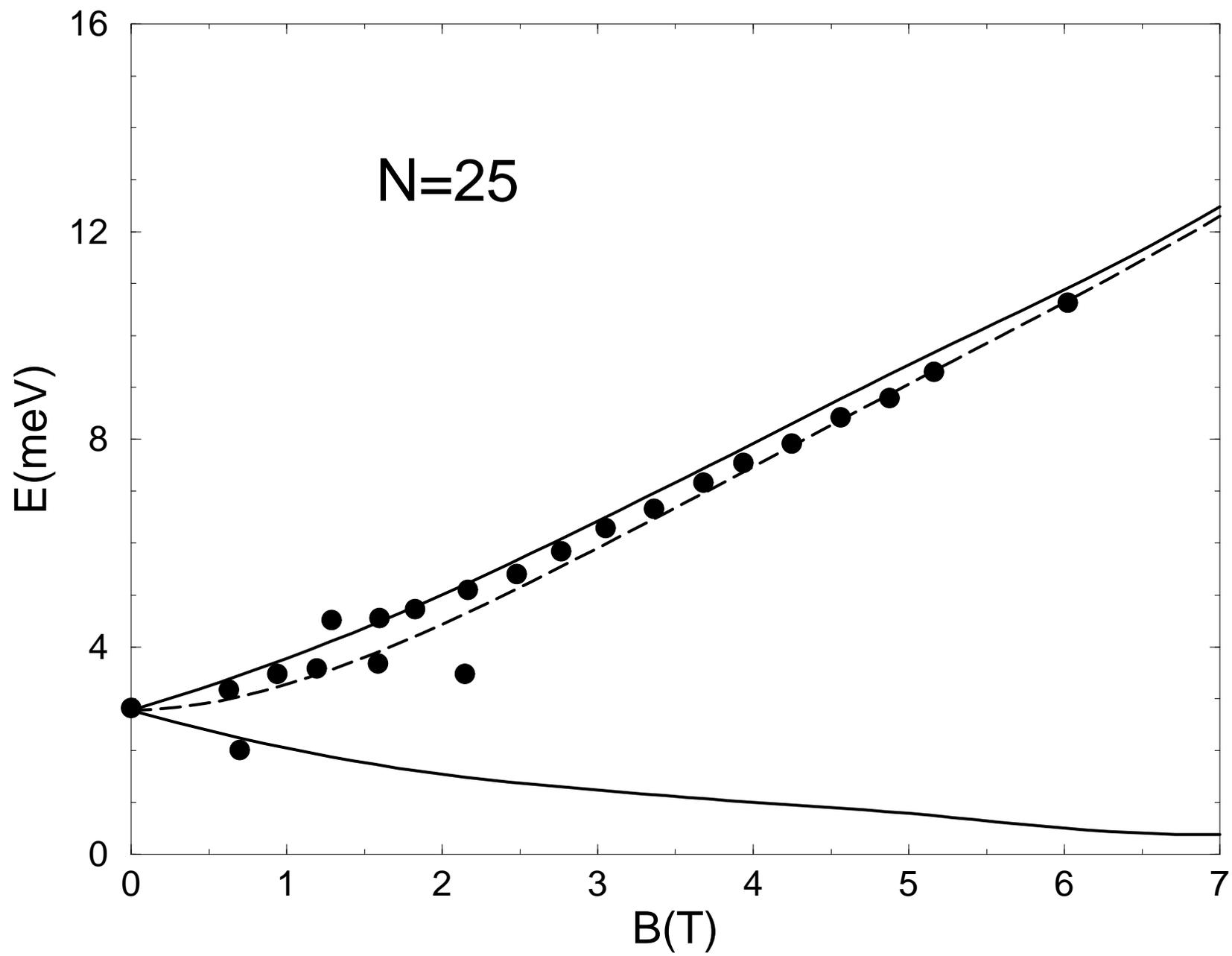
\end{document}